\documentclass[runningheads]{llncs}
\usepackage[T1]{fontenc}
\usepackage{makecell}
\usepackage{booktabs,amsmath,microtype,xcolor,tikz}
\usetikzlibrary{positioning,arrows.meta}
\newcommand{\mumap}{$\mu$-map}

\begin{document}
\title{Topogram-Gated Multi-Modal Pseudo-CT Synthesis for PET/MR Attenuation Correction}
\titlerunning{Topogram-Gated Pseudo-CT Synthesis for PET/MR}
\author{Joris Wuts\inst{1,2,3,4} \and
Jakub Ceranka\inst{1,2} \and
Vicky De Ridder\inst{6} \and
Jef Vandemeulebroucke\inst{1,2,5}}
\authorrunning{J. Wuts et al.}
\institute{Department of Electronics and Informatics (ETRO.RDI), Vrije Universiteit Brussel (VUB), Pleinlaan 2, 1050 Brussels, Belgium \and
imec, Kapeldreef 75, 3001 Leuven, Belgium \and
Research Foundation - Flanders (FWO), Leuvenseweg 38, 1000 Brussels, Belgium \and
Institute for Experimental and Clinical Research (IREC), Université catholique de Louvain (UCLouvain), Avenue Hippocrate 55, 1200 Woluwe-Saint-Lambert, Belgium \and
Department of Radiology, Universitair Ziekenhuis Brussel (UZ Brussel), Vrije Universiteit Brussel (VUB), Laarbeeklaan 101, 1090 Brussels, Belgium \and
Nuclivision BV, Edward Pynaertkaai 87, 9000 Gent, Belgium}
\maketitle
\let\thefootnote\relax\footnotetext{Corresponding author: \email{Joris.Sebastiaan.Wuts@vub.be}}
\begin{abstract}
As part of the Big Cross-Modal Attenuation Correction Challenge (BIC-MAC), we aimed to synthesise pseudo-CT images for PET/MR attenuation correction from whole-body non-attenuation-corrected PET (NAC-PET), Dixon MRI, and a two-dimensional topogram image. We developed a residual 3D U-Net in which a dedicated 2D topogram encoder provides bounded, multi-scale feature gating. The network was supervised jointly in CT and 511-keV attenuation-map space. Two complementary models, one emphasizing activity and cranial regions and one including a projection-domain attenuation-correction-factor regulariser, were combined by equal averaging in CT space. In controlled development experiments, the two training objectives yielded complementary error profiles. On the online BIC-MAC validation leaderboard, the ensemble achieved a \mumap{} MAE of 0.005656, whole-body SUV MAE of 0.0356, organ bias of 2.65\%, and brain outlier score of 0.0294. Topogram-gated multi-modal learning combines local volumetric information with global projected anatomy and provides a practical, physically informed approach to pseudo-CT synthesis for PET/MR.
\end{abstract}

\section{Introduction}
Positron emission tomography (PET) quantification relies on a map of 511-keV linear attenuation coefficients to compensate for photon loss along each line of response. Computed tomography (CT) provides a practical proxy for this map in PET/CT, but PET/MR must infer attenuation from MR acquisitions that principally reflect proton properties rather than electron density. In particular, air and cortical bone can both exhibit little MR signal despite having markedly different attenuation. Conventional MR attenuation-correction (MRAC) methods based on tissue classes or registered atlases consequently leave appreciable quantitative error \cite{hofmann2011}. Deep learning enables continuous pseudo-CT synthesis and has reduced PET reconstruction error relative to such MR-only baselines \cite{liu2018}.

The BIC-MAC dataset supplies three complementary observations: a volumetric Non-Attenuation Corrected PET (NAC-PET) acquisition, a volumetric Dixon MRI acquisition, and a single two-dimensional topogram. PET provides functional and body-contour information, MR resolves soft-tissue anatomy, and the topogram exposes long-range projected skeletal structure. The latter is intrinsically two-dimensional and therefore cannot be naively concatenated with a 3D image volume without imposing an artificial depth representation.

In our approach, we propose a topogram-gated residual 3D U-Net that integrates these modalities while preserving their native dimensionality. Our contributions are three-fold: (i) multi-scale, centred topogram gating of a volumetric image encoder; (ii) joint CT and attenuation-map supervision augmented by transverse projection consistency; and (iii) an ensemble of complementary models that can be evaluated sequentially within a single-model memory footprint.

\section{Methods}
\subsection{Challenge dataset and model input}
The provided dataset consisted of 75 labelled examinations. Each examination contained a rigidly co-registered NAC-PET acquisition and reference CT volume acquired during the same PET/CT scan session, together with a Dixon MRI acquisition obtained from a separate MR session. The multimodal inputs therefore contain inherent registration challenges, particularly between MRI and the PET/CT-derived volumes, due to differences in acquisition time, patient positioning, and scanner geometry. In addition, each examination included a two-dimensional topogram and the body mask supplied by the challenge. Our model used the following data:
\begin{itemize}
    \item The in-phase whole-body DIXON MRI image.
    \item NAC-PET image defining the spatial reference grid used for model inputs and pseudo-CT output.
    \item The topogram representing a projection in the XZ plane and thus lacking the orthogonal Y dimension.
\end{itemize} 
This combination provided functional, anatomical, and global projection information while retaining CT image as the spatial reference for attenuation-related synthesis.

\subsection{Preprocessing and target representation}
For each subject, PET is clipped to 3000 intensity units, and log-compressed as $\log(1+I)$, where $I$ is the intensity value at any voxel location. MRI is bi-linearly resampled to the NAC-PET grid with border padding and normalised over non-zero voxels.

The network predicts a continuous scaled CT image $\mathbf{s}$. At output time, it is transformed to Hounsfield units (HU) as $\mathbf{h}=3000\mathbf{s}-1000$. For physically meaningful supervision, HU is converted to a 511-keV \mumap{} with the fixed bilinear transform
\begin{equation}
\mu(\mathbf{h})=\begin{cases}
9.6\times10^{-5}(\mathbf{h}+1000), & \mathbf{h}<47,\\
5.10\times10^{-5}(\mathbf{h}+1000)+4.71\times10^{-2}, & \mathbf{h}\geq47,
\end{cases}
\label{eq:hu-mu}
\end{equation}
where values are in cm$^{-1}$. This allows CT-compatible challenge output while directly penalising the quantity that determines attenuation correction.

\subsection{Topogram-gated residual 3D U-Net}
Our architecture is a residual volumetric U-Net \cite{cicek2016} with two input channels and five levels $[32,64,128,256,512]$. Each stage has one residual block with two $3\times3\times3$ convolutions, instance normalisation, and leaky ReLU; the bottleneck uses 0.2 dropout. Max-pooling and transposed convolutions implement down- and upsampling.

\begin{figure}[t]
\centering
\resizebox{\linewidth}{!}{%
\begin{tikzpicture}[
  font=\scriptsize,
  >=Latex,
  input/.style={draw,rounded corners=2pt,align=center,minimum height=8mm,minimum width=20mm,fill=gray!8},
  path3d/.style={draw,rounded corners=2pt,align=center,minimum height=10mm,minimum width=25mm,fill=gray!12},
  path2d/.style={draw,rounded corners=2pt,align=center,minimum height=10mm,minimum width=25mm,fill=orange!14},
  outputbox/.style={draw,rounded corners=2pt,align=center,minimum height=10mm,minimum width=24mm,fill=gray!12},
  arr/.style={->,line width=.6pt},
  modarr/.style={->,line width=.75pt,draw=orange!70!black}
]

\node[input] (pet) at (0,0.9) {NAC-PET\\$\log(1+I)$};
\node[input] (mr) at (0,-0.3) {Dixon MRI\\in-phase used};
\node[path2d] (topogram) at (0,-2.0) {2D topogram\\XZ projection};

\node[path3d] (enc) at (3.2,0.3) {3D encoder};
\node[path3d] (bottleneck) at (6.8,0.3) {Bottleneck};
\node[path3d] (dec) at (10.4,0.3) {3D decoder};
\node[outputbox] (ct) at (14.0,0.3) {Predicted\\pseudo-CT};

\node[path2d] (topenc) at (3.4,-2.0) {2D topogram\\encoder};

\draw[arr] (pet.east) -- ++(0.45,0) |- (enc.west);
\draw[arr] (mr.east) -- ++(0.45,0) |- (enc.west);
\draw[arr] (enc.east) -- (bottleneck.west);
\draw[arr] (bottleneck.east) -- (dec.west);
\draw[arr] (dec.east) -- (ct.west);

\draw[arr] (topogram.east) -- (topenc.west);

\draw[modarr] (topenc.north) -- ++(0,0.55) -| (enc.south);
\draw[modarr] (topenc.east) -- ++(1.3,0) -| (bottleneck.south);

\node[font=\scriptsize,orange!70!black,below=1mm of enc] {centred gating};
\node[font=\scriptsize,orange!70!black,below=1mm of bottleneck] {centred gating};

\end{tikzpicture}%
}
\caption{Architecture of the proposed multimodal pseudo-CT synthesis network. NAC-PET and the selected in-phase Dixon MRI volume form the two input channels of a 3D residual U-Net comprising an encoder, bottleneck, and decoder. The two-dimensional topogram is encoded separately in the XZ plane, shown in yellow, and provides centred gating to both the 3D encoder and bottleneck before pseudo-CT prediction.}
\label{fig:method}
\end{figure}
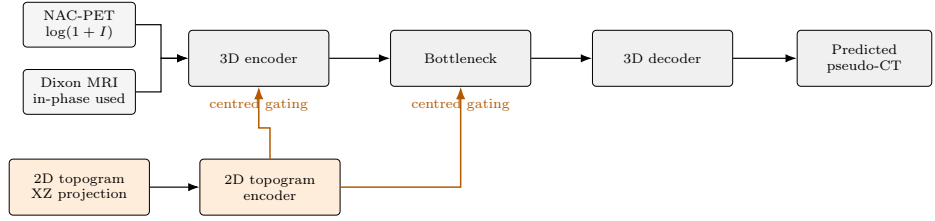

The topogram is processed by a 2D encoder with channels $[16,32,64]$. Features at encoder levels two and three and at the volumetric bottleneck are bilinearly resized in the observed XZ plane, repeated over the missing Y dimension, and mapped to the number of 3D feature channels by a $1\times1\times1$ convolution. They form a centred multiplicative gate:
\begin{equation}
 \widetilde{F}=F\odot\left[1+\gamma\tanh\bigl(g(T)\bigr)\right], \qquad \gamma=0.25,
\label{eq:gate}
\end{equation}
where $F$ is a volumetric feature map, $T$ is an expanded topogram feature map, and $g$ is the learned projection. Centering constrains the topogram-free behaviour close to the volume stream while enabling the 2D projection to modulate relevant anatomical regions (Fig.~\ref{fig:method}).

\subsection{Training objective and complementary models}
\begin{figure}[t]
\centering
\resizebox{0.92\linewidth}{!}{%
\begin{tikzpicture}[
  font=\scriptsize,
  >=Latex,
  baseloss/.style={
    draw,
    rounded corners=2pt,
    align=center,
    text width=58mm,
    minimum height=14mm,
    inner sep=5pt,
    fill=gray!12
  },
  modelone/.style={
    draw,
    rounded corners=2pt,
    align=center,
    text width=58mm,
    minimum height=20mm,
    inner sep=5pt,
    fill=blue!12,
    draw=blue!55!black
  },
  modeltwo/.style={
    draw,
    rounded corners=2pt,
    align=center,
    text width=58mm,
    minimum height=20mm,
    inner sep=5pt,
    fill=red!10,
    draw=red!55!black
  },
  bluearr/.style={
    ->,
    line width=.8pt,
    draw=blue!60!black
  },
  redarr/.style={
    ->,
    line width=.8pt,
    draw=red!60!black
  }
]

\node[baseloss] (base) at (0,0) {
\textbf{Masked base loss}\\[1mm]
$\mathcal{L}_{\mathrm{base}}
=
\mathcal{L}_{\mathrm{CT-L1}}
+
0.75\,\mathcal{L}_{\mu\mathrm{-L1}}$\\
Computed within the supplied prediction mask
};

\node[modelone] (m1) at (-3.8,-3.4) {
\textbf{Model 1: activity + cranial emphasis}\\[1mm]
Weighted masked $\mathcal{L}_{\mathrm{base}}$\\
Head/skull weighting + NAC-PET activity weighting
};

\node[modeltwo] (m2) at (3.8,-3.4) {
\textbf{Model 2: projection consistency}\\[1mm]
$\mathcal{L}_{\mathrm{base}}
+
0.0025\,\mathcal{L}_{\mathrm{ACF}}$\\
Transverse ACF projection-consistency penalty
};

\draw[bluearr]
(base.south west) -- ++(0,-0.7) -| (m1.north);

\draw[redarr]
(base.south east) -- ++(0,-0.7) -| (m2.north);

\end{tikzpicture}%
}

\caption{
Training objectives for the two complementary ensemble members. Both models start from the same masked base objective combining CT-space and 511-keV $\mu$-map supervision. Model~1 (blue) applies activity- and cranial-region weighting to this base loss, whereas Model~2 (red) augments it with the transverse ACF projection-consistency term.
}
\label{fig:losses}
\end{figure}
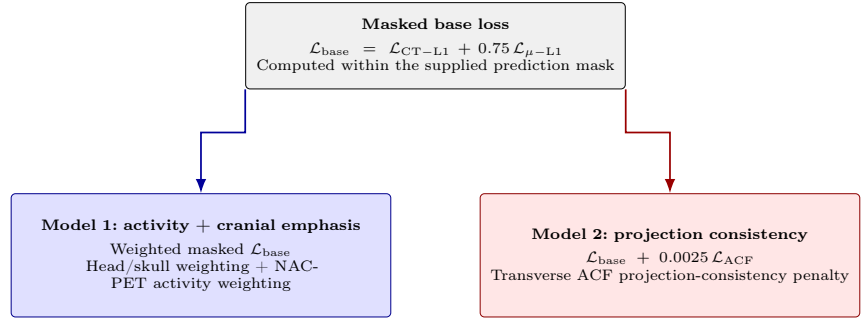

All models minimise a masked base objective,
\begin{equation}\mathcal{L}_{\mathrm{base}}=\mathcal{L}_{\mathrm{CT-L1}}+0.75\mathcal{L}_{\mu\mathrm{-L1}},\label{eq:base-loss}\end{equation}
over the supplied prediction mask. The first term compares scaled CT; the second is L1 error after Eq.~\ref{eq:hu-mu}. The two ensemble members use the same backbone, data representation, and optimiser, but differ in their training emphasis (Fig~\ref{fig:losses}).

\paragraph{Activity- and cranial-emphasis model.} Head-centred patches are sampled with probability 0.25, with head and skull weights of 1.1 and 1.5, respectively. Voxel weighting further incorporates $1+(\mathrm{clip}(I/8,0,1))^{0.5}$, where $I$ denotes NAC-PET intensity. MRI-only translation translations are applied with probability 0.75, up to four voxels per axis.

\paragraph{Projection-consistency model.} The second model uses unweighted base sampling and adds a transverse attenuation-correction-factor (ACF) term with weight 0.0025. Inspired by previous physics-informed projection losses for PET attenuation map generation \cite{shi2019}, this term introduces projection-domain consistency in addition to voxel-wise supervision. Let $M$ denote the supplied prediction mask and let $\Delta_d(r)$ represent the masked integral of the difference between predicted and reference \mumap{} along a transverse ray $r$ in direction $d\in\{x,y\}$, expressed in centimetres:
\[
\Delta_d(r)=\int_{\mathrm{ray}\ r} M(\mathbf{x})\left(\mu_{\mathrm{pred}}(\mathbf{x})-\mu_{\mathrm{ref}}(\mathbf{x})\right)\,dl .
\]
For each direction $d$, let $\mathcal{R}_d$ denote the set of axis-aligned transverse rays, with $r\in\mathcal{R}_d$ indexing an individual ray. The additional term is
\begin{equation}
\mathcal{L}_{\mathrm{ACF}}=\frac{1}{2}\sum_{d\in\{x,y\}}\operatorname*{mean}_{r\in\mathcal{R}_d}\left|\exp\left(\operatorname{clip}(\Delta_d(r),-1,1)\right)-1\right|.
\label{eq:acf-loss}
\end{equation}
Whereas previous projection-based losses compare line-integral attenuation values directly \cite{shi2019}, this formulation evaluates the corresponding exponential attenuation correction factors, which more closely represent the multiplicative correction applied during PET attenuation correction. The term therefore penalises errors in the attenuation correction factor accumulated along axis-aligned transverse rays, complementing the local voxelwise losses, while remaining computationally efficient and differentiable for network optimisation.

Training uses $192^3$ patches, four patches per GPU, two H200 GPUs, FP32 AdamW \cite{kingma2015} (learning rate $10^{-4}$, $(\beta_1,\beta_2)=(0.9,0.999)$, and weight decay $10^{-5}$), gradient clipping at 1.0, and a constant learning-rate schedule. Final models were trained using all labelled examinations; hence, development experiments described below are used for model selection rather than for an unbiased estimate of final ensemble performance.

\subsection{Inference and ensembling}
At inference, each model receives the complete topogram for every image window. We use Gaussian-weighted sliding-window inference with 192$^3$ windows, 50\% overlap, and a window batch size of one, implemented with MONAI \cite{cardoso2022}. We convert each scaled prediction to float32 HU, then average in CT space:
\begin{equation}h_{\mathrm{ens}}=\tfrac{1}{2}h_{\mathrm{activity/head}}+\tfrac{1}{2}h_{\mathrm{ACF}}.\end{equation}
Models are loaded and evaluated sequentially, so the ensemble retains approximately one-model peak GPU memory. The output is a float32 NIfTI pseudo-CT with the NAC-PET affine.

\section{Results}
We performed controlled single-model development experiments to identify complementary training objectives. Activity- and cranial-region emphasis achieved the lowest brain-outlier metric (0.012779). Projection-consistency regularisation achieved the lowest whole-body SUV MAE (0.039117), organ-bias score (2.8817), and \mumap{} MAE (0.005803) among the evaluated variants (Table~\ref{tab:results}). These complementary results motivated their equal-weight ensemble.

\begin{table}[t]
\caption{Development and final ensemble results (lower is better). The ensemble corresponds to the submitted BIC-MAC validation result.}
\label{tab:results}
\centering
\scriptsize
\setlength{\tabcolsep}{9pt} 
\begin{tabular}{lrrrr}
\toprule
Model 
& \makecell{\mumap{}\\MAE}
& \makecell{SUV\\MAE}
& \makecell{Organ bias\\(\%)}
& \makecell{Brain\\outlier}\\
\midrule
Activity + cranial emphasis 
& 0.005848 & 0.039490 & 3.02 & \textbf{0.012779}\\
Projection consistency ($\lambda=0.0025$) 
& 0.005803 & 0.039117 & 2.89 & 0.028453\\
Ensemble 
& \textbf{0.005656} & \textbf{0.0356} & \textbf{2.65} & 0.0294\\
\bottomrule
\end{tabular}
\end{table}
\section{Discussion \& Conclusion}
The proposed topogram-gated framework integrates complementary anatomical and functional information while respecting the different dimensionality of the input modalities. The topogram provides long-range projected anatomical context through bounded feature modulation, while the spatial localisation of the predicted pseudo-CT remains determined by the volumetric PET and MRI representations. This allows the model to exploit projection-level information without introducing an explicit depth reconstruction step.

The two training objectives encourage complementary properties of the predicted attenuation maps. Activity and cranial weighting prioritise regions with high quantitative relevance and challenging anatomy, whereas ACF regularisation constrains accumulated transverse attenuation differences. Equal-weight averaging in CT space provides a simple and stable ensemble strategy without requiring additional optimisation on the limited development data.

The ACF loss provides a computationally efficient approximation of projection-domain attenuation consistency, but it does not model the complete PET acquisition geometry or reconstruction process. Future work could investigate scanner-specific differentiable projectors, cross-validated ensemble weighting, and uncertainty estimation to further improve robustness and quantitative reliability.

In this work, we presented a topogram-gated residual 3D U-Net for pseudo-CT synthesis in PET/MR as part of the BIC-MAC MICCAI 2026 challenge submission. The proposed ensemble combines volumetric PET and MRI information with 2D projection context, uses supervision in both CT and attenuation-map spaces, and enables reproducible inference within a limited memory footprint. This approach provides a physically informed framework for PET/MR attenuation-map synthesis.

\section*{Acknowledgments}
The resources and services used in this work were provided by the VSC (Flemish Supercomputer Center), funded by the Research Foundation – Flanders (FWO) and the Flemish Government. This study was funded by the Research Foundation – Flanders (FWO), grant number 1S43623N.

\section*{Declaration of generative AI and AI-assisted technologies in the writing process}
During the preparation of this work, the author(s) used ChatGPT (GPT-5, OpenAI) to assist with formatting tables and figures and to improve academic clarity. After using this tool, the author(s) reviewed and edited all content and take full responsibility for the content of the published article.

\bibliographystyle{unsrt}
{\small\bibliography{reference}}
\end{document}